\documentclass[aps, prl, twocolumn,superscriptaddress,amsmath,amssymb,floatfix]{revtex4-1}

\usepackage{graphicx}
\usepackage[usenames]{color} 
\usepackage{bm}
\usepackage{amsmath}

\newcommand{\be}{\begin{equation}} 
\newcommand{\ee}{\end{equation}}
\newcommand{\bea}{\begin{eqnarray}} 
\newcommand{\eea}{\end{eqnarray}}

\begin{document}

\title{Quantifying density fluctuations in water at a hydrophobic surface: evidence for critical drying}

\author{Robert Evans}  
\affiliation{H. H. Wills Physics Laboratory, University of Bristol, Royal Fort, Bristol BS8 1TL, United Kingdom}
\author{Nigel B. Wilding} 
\affiliation{Department of Physics, University of Bath, Bath BA2 7AY,
United Kingdom} 

\begin{abstract}

Employing smart Monte Carlo sampling techniques within the grand canonical ensemble, we
investigate the properties of water at a model hydrophobic substrate. By reducing the strength of
substrate-water attraction we find that fluctuations in the local number density, quantified by a
rigorous definition of the local compressibility $\chi(z)$, increase rapidly for
distances $z$ within $1$ or $2$ molecular diameters from the substrate as the degree of hydrophobicity,
measured by the macroscopic contact angle $\theta$, increases. Our simulations provide evidence for a
continuous (critical) drying transition as the substrate-water interaction becomes very weak: $\cos(\theta)\to -1$. We
speculate that the existence of such a transition might account for earlier simulation observations
of strongly enhanced density fluctuations.

\end{abstract}

\maketitle

All physical scientists would agree that for water at a flat substrate
a contact angle $\theta>90^\circ$ defines the substrate as
hydrophobic. For such values Young's equation implies that the
substrate-vapor interfacial tension is lower than the substrate-liquid
tension, i.e. the substrate prefers vapor to liquid. $\theta$ is
determined by surface chemistry, which in turn determines the strength
and range of substrate-fluid interactions.   A key question, much
discussed by the chemical physics communities, is whether there is an
effective indicator of local ordering of water, manifest at {\em
  microscopic} distances from the substrate, which might correlate
with the degree of hydrophobicity as measured by the {\em macroscopic}
thermodynamic quantity $\theta$. Quantifying the character and spatial
extent of water ordering at hydrophobic entities is important across
several disciplines, ranging from applied physics and materials
science, where understanding slip lengths of water, shown to be correlated with substrate hydrophobicity, is crucial in
microfluidics \cite{Bocquet:2010fk}, to bio-physical processes such as
protein-folding and micelle and membrane formation
\cite{Ball:2008rz}. Many studies have focussed on the average one-body
density, i.e. the density profile of oxygen atoms near the
substrate. Experimentally this is difficult to measure and several
conflicting results were reported. Nevertheless, by 2009 a consensus
emerged from x-ray reflectivity measurements that for water near a
variety of hydrophobic self-assembled-monolayers (SAMs), there is a
region of density depletion corresponding to only a fraction of a
water monolayer \cite{Mezger:2006zl, Ocko:2008fv, Mezger:2010lq}. This
was challenged by Chattopadhyay et.al. \cite{Chattopadhyay:2010aa},
who reported larger depletion lengths, increasing with $\theta$, for
water at fluoroalkylsilane SAMs. For the largest contact angle
$\theta=120^\circ$ the depletion length was $8$\AA, corresponding to
about three water layers. However, in a re-analysis
\cite{Mezger:2011aa} of the x-ray reflectivity data of
Ref.~\cite{Chattopadhyay:2010aa} it was argued that such a large
hydrophobic gap is likely to be an artefact of the data analysis that
was used. This is disputed \cite{Chattopadhyay:2011aa}. Certainly such
large depletion lengths are at odds with results of many computer
simulations of water models, where the thickness of the depleted
density region varies typically between $1.5-2.0$\AA~for $\theta$
between about $110^\circ$ and $130^\circ$ \cite{Janecek:2007aa}.

Other simulation studies have focussed on the density fluctuations of
water at model hydrophobic substrates, arguing that {\em some} measure
of the local compressibility might provide a better indicator of
hydrophobicity than does the local density profile. The group of Garde
\cite{Acharya:2010aa} takes this stance and the review by Jamadagni
et.~al. \cite{Godawat:2011aa} places their work in the context of
bio-molecular systems. Chandler and co-workers \cite{Chandler:2007aa,
  Patel:2010dz,Willard:2014aa} and Mittal and Hummer
\cite{Mittal:2010aa} also emphasize that density fluctuations can be
enhanced at hydrophobic substrates. Although these studies identify
some underlying phenomenology, the various measures of the local
compressibility introduced are ad-hoc. In particular these do not
correspond to integrals over density correlation functions in the
inhomogeneous liquid. Clear insights into the nature of the underlying
fluctuations requires such a measure \cite{Evans:SM}. Choice of
ensemble is important. Most members of the community simulating water,
choose not to work grand canonically. In most real interfacial
situations there is a reservoir and it is natural to vary the chemical
potential $\mu$ of the liquid. Following a recent analysis
\cite{Evans:2015aa} of density fluctuations in a Lennard-Jones (LJ)
liquid at a planar substrate (wall), we argue that the most
appropriate definition \cite{Evans:SM} of the local compressibility,
for a fixed confining volume, is
\be
\chi(z)\equiv(\partial \rho(z)/\partial \mu)_T\:,
\label{eq:chi}
\ee
where $\rho(z)$ is the average one-body density,~$z$ is the distance normal to the substrate, and the temperature $T$ is fixed. For a bulk fluid with constant density $\rho_b$,
$\chi(z)\equiv\chi_b=\rho_b^2\kappa_T$, where $\kappa_T$ is the isothermal compressibility. The definition (\ref{eq:chi}) is consistent with $\chi(z)$ as an integral over the density-density
correlation function {\em and} yields a unique fluctuation formula for this quantity; See Eqs.~(8,9) of \cite{Evans:SM}. In ref.~\cite{Evans:2015aa} it was
shown using classical density functional theory (DFT) that for $z$ close to the substrate $\chi(z)/\chi_b$ becomes large as the substrate becomes more solvophobic, i.e. $\theta$ increases.
The effects are not small: for $\theta\sim 160^\circ$ the ratio is about $25$.

In this Letter we determine the local compressibility of the extended simple point charge (SPC/E)
model of water \cite{Berendsen:1987aa} near a planar substrate. On reducing the strength of
substrate-water attraction, thereby increasing the hydrophobicity as measured by $\theta$, we find
$\chi(z)$ increases in a similar fashion to the LJ case \cite{Evans:2015aa}. We focus on the
approach to complete drying, $\cos(\theta)\to -1$ at bulk vapor-liquid coexistence $\mu=\mu_{cx}$.
Determining the nature of this transition continues to be challenging, even for simple fluids at
solvophobic substrates. Using smart sampling techniques within Grand Canonical Monte Carlo (GCMC)
simulations we find evidence for a critical drying transition in SPC/E water at a weakly attractive
substrate. We note that recent MD simulations \cite{Willard:2014aa,Godawat:aa} of the same water
model also investigate very weak substrate-water attraction. Although both studies attempt to link
the growth of density fluctuations to increasing $\theta$, neither simulation measures $\theta$ accurately
so proximity to the drying point is uncertain. Moreover, neither makes explicit the possibility of a
critical drying transition. We argue the latter is key to understanding the properties of water
models in the extreme hydrophobic regime and, in particular, the enhanced density fluctuations that
are observed.

We choose SPC/E as a simple but realistic model of water, known to provide a reasonable account of
bulk vapor-liquid coexistence and the vapor-liquid surface tension of real water
\cite{Kumar:2013aa,Vega:2007aa}. Moreover, this model has often been employed in simulations of
water at hydrophobic substrates; for examples of MD studies see refs.
\cite{Willard:2014aa,Godawat:aa,Acharya:2010aa,Giovambattista:2006sf} and for GCMC studies see
refs.~\cite{Kumar:2013aa,Kumar:2013kx,Bratko:2007aa,Bratko:2010aa}. Kumar and Errington employ
simulation techniques similar to the present but they do not measure $\chi(z)$, i.e. they do not
access {\em local} density fluctuations. Rather they measure \cite{Kumar:2013kx} the surface excess
compressibility $\chi_{ex}$ which is an {\em integrated} measure of `excess' density fluctuations
throughout the system \cite{Evans:2015aa,Evans:1987aa,Evans:SM}

Our simulations were performed at $T=298 {\rm K}$. Since liquid water
at room temperature is too dense for standard GCMC to operate
effectively, smart sampling techniques were implemented.
Configurational Bias MC was used to insert, delete, translate and rotate
molecules \cite{Martin:2013zl}, while Transition Matrix MC
\cite{Smith:1995kx}, Multicanonical Sampling \cite{berg1992} and
Histogram Reweighting \cite{Ferrenberg1988} were deployed to smoothly
connect the vapor and liquid regions of configuration space.  Together
these methods allowed us to simulate modest system sizes of order a
few hundred water molecules very accurately. Although MD simulations
typically deal with greater numbers of molecules, the GCE is the
appropriate ensemble for accurately studying $\chi(z)$ and thermodynamic quantities such as $\cos(\theta)$ and surface phase behaviour
because it permits direct study of the fluctuations which characterise
phase transitions. As these fluctuations occur on the scale of the
system itself, the GCE is less afflicted by finite-size effects than
other simulation ensembles.

We consider two simulation setups: (i) a fully periodic cubic box of
volume $V=L^3$; (ii) a semi-periodic cuboidal slit geometry of volume
$V=L^2D$, in which the oxygen atoms interact with a pair of
symmetry-breaking walls separated by distance $D$. For the latter
geometry, the single wall-oxygen potential takes the form $V_s(z) = \alpha
\epsilon_{wf}[ 2/15 (\sigma_{wf}/z)^9 - (\sigma_{wf}/z)^3 ]$, where
$\epsilon_{wf}$ is the wall-fluid interaction strength and $z$ is the
distance from the wall. $\alpha$ is a constant, the choice which sets the units of energy (see Supplementary Material \cite{Evans:SM}),  while $\sigma_{wf}$ sets the length scale for
wall-fluid interactions which we assigned to be $3.5$\AA, in
accordance with previous studies \cite{Kumar:2013kx}.

The choice of $\epsilon_{wf}$ controls the degree of hydrophobicity,
and hence the contact angle $\theta$. Our use of the GCE permits us
to calculate $\theta$ directly from Young's equation,
\be
\gamma_{vl}\cos(\theta) = \gamma_{wv}-\gamma_{wl}\:.
\label{eq:Young}
\ee 
Here $\gamma_{vl}$ is the vapor-liquid interfacial tension and
$\gamma_{wv}$ and $\gamma_{wl}$ are the wall-vapor and wall-liquid
interfacial tensions, respectively. Provided one can calculate $\gamma_{vl}$ and
$\gamma_{wv}-\gamma_{wl}$ at
$\mu=\mu_{cx}$, $\cos(\theta)$ is given. Both quantities are directly obtainable from
measurements of the probability distribution $p(\rho)$ of the
fluctuating molecular number density $\rho=N/V$ in the appropriate
simulation geometry.  Specifically, studies of $p(\rho)$ in the
fully periodic system allow estimates of both $\mu_{cx}$ and $\gamma_{vl}$,
while studies of $p(\rho)$ in the slit geometry allow measurements of
$\gamma_{wv}-\gamma_{wl}$.

For the periodic system at vapor-liquid coexistence, $p(\rho)$
exhibits a pair of equally weighted peaks \cite{Borgs1992} separated
by a flat probability `valley'. The low and high density peaks
correspond to pure vapor and liquid phase states respectively, while
the flat valley corresponds to liquid-slab configurations in which a
pair of liquid-vapor interfaces align parallel to one face of the
simulation box. Since in the probability valley $p_{\rm min}$ is
typically many decades smaller than the peak probabilities $p_{\rm
  vap}$ and $p_{\rm liq}$, standard sampling cannot plumb the valley
depth.  For this reason, Transition Matrix and Multicanoncial MC
techniques \cite{Smith:1995kx} were used to accumulate $p(\rho)$ in
histogram form across the full range of density from vapor to
liquid. Initially $p(\rho)$ was determined for a near-coexistence
state point. The distribution was then reweighted
\cite{Ferrenberg1988} with respect to $\mu$ to determine $\mu_{cx}$
via the equal peak weight criterion. Once obtained, the coexistence
form of $p(\rho)$ permits an estimate of $\gamma_{vl}$ 
\cite{Binder82,Errington2003}:
\be 
\gamma_{vl}=(2\beta L^2)^{-1}\ln(p_{\rm max}/p_{\rm min})\:,  
\label{eq:gamlv}
\ee 
where $\beta=(k_BT)^{-1}$ and $p_{\rm max}\equiv p_{\rm vap}=p_{\rm liq}$ is the height of the pure phase peaks.

\begin{figure}[h]
\includegraphics[type=pdf,ext=.pdf,read=.pdf,width=0.95\columnwidth,clip=true]{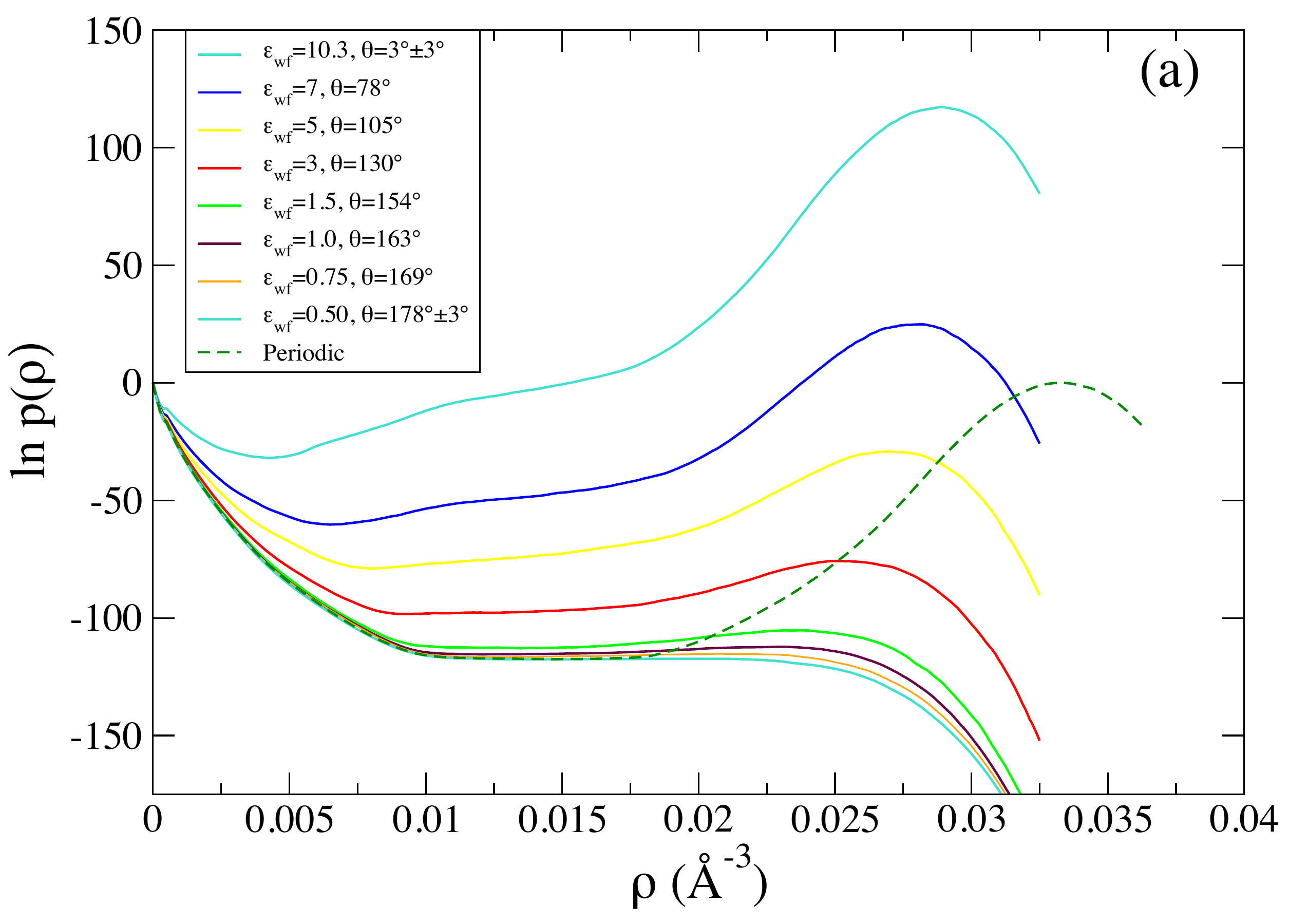}
\includegraphics[type=pdf,ext=.pdf,read=.pdf,width=0.95\columnwidth,clip=true]{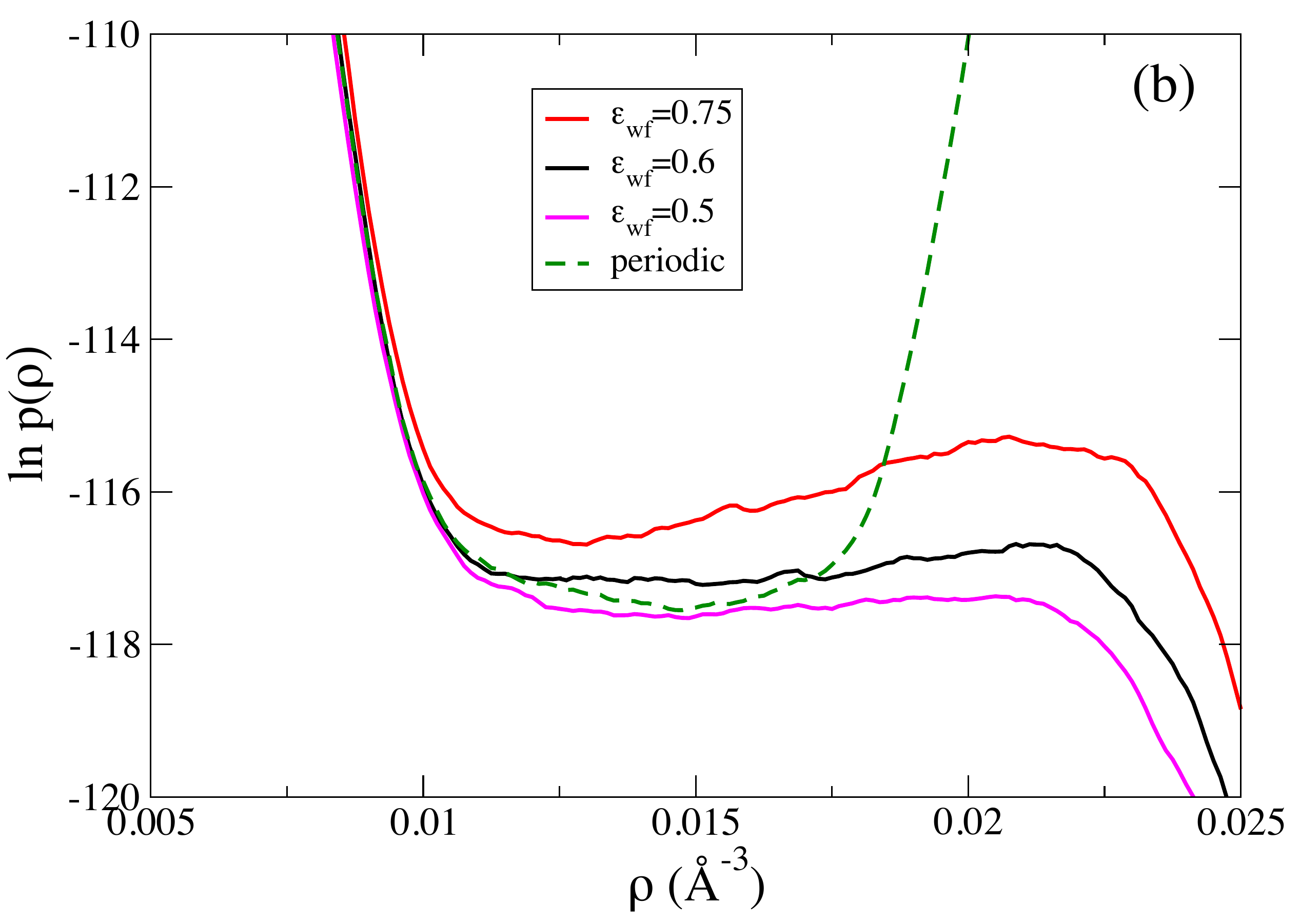}
\vspace*{-2mm}
\caption{(Color online). {\bf (a)} The measured forms of $p(\rho)$ for a selection of
  values of $\epsilon_{wf}$ spanning the range from wetting to drying;
  the contact angles $\theta$ are shown. The system size is
  $L=D=20$\AA.  {\bf (b)} A closeup of the region close to the drying
  transition. Note the continuous erosion of the liquid peak as
  $\epsilon_{wf}$ decreases. In {\bf (a)} and {\bf (b)} the dashed
  line is $p(\rho)$ for the periodic system with
  $L=20$\AA.}
  \label{fig:prhoslit}
\end{figure}

In a similar manner, estimates of $p(\rho)$ at coexistence were
accumulated for the slit geometry at various $\epsilon_{wf}$. The
resulting histograms (Fig.~\ref{fig:prhoslit}) typically exhibit two
peaks --one at low density corresponding to wall-vapor (i.e. vapor at the wall) configurations,
and another at high density corresponding to wall-liquid (liquid at the wall)
configurations \footnote{At $T=298K$ the vapor peak occurs at a density which
is so low it corresponds to an average of less than one molecule in
our simulation box. Accordingly the peak maximum appears at zero
density in the histogram.}. For a given $\epsilon_{wf}$,
$\gamma_{wv}-\gamma_{wl}$ is calculated from the measured ratio of
vapor and liquid peak heights in $p(\rho)$ \cite{Muller:2000fv}:
\be
\gamma_{wv}-\gamma_{wl}=-(2\beta L^2)^{-1}\ln(p_{\rm vap}/p_{\rm liq})\:.
\label{eq:gamdiff}
\ee

We now discuss the pertinent features of Fig.~\ref{fig:prhoslit}, noting that $-\ln p(\rho)$ measures the grand potential of the fluid.  For
large values of $\epsilon_{wf}$, wall-liquid configurations are much
more probable than wall-vapor configurations.  Indeed for
$\epsilon_{wf}\approx 10.3$, we find $\cos(\theta)=1$, corresponding
to the transition to complete wetting by liquid water. Reducing
$\epsilon_{wf}$ below this value takes the system first into the
partially wet (hydrophilic) regime for which $0^\circ<\theta<90^\circ$. Thereafter,
for $\epsilon_{wf}\lesssim 6$, the system enters the 
partially dry (hydrophobic) regime in which $90^\circ<\theta<180^\circ$ and where
wall-vapor configurations are favoured over wall-liquid ones \footnote{For the
slit system that we consider, capillary
evaporation should occur in the partially dry regime $90^\circ<\theta<180^\circ$. Nevertheless it is possible to study the
properties of the wall-liquid system because the liquid peak in
$p(\rho)$ persists metastably; the free energy barrier is enormous.}.

As $\epsilon_{wf}$ is reduced within the hydrophobic regime,
$\epsilon_{wf}<6$, Fig.~\ref{fig:prhoslit} shows that the height of
the liquid peak diminishes progressively, until it vanishes smoothly
into a plateau in $\ln p(\rho)$ at a low wall strength which, for this
system size, is $\epsilon_{wf}=0.5(1)$. Interestingly, the vanishing
of the liquid peak occurs precisely at the drying point $\cos(\theta)=
-1$. This is clear from the dashed curve in Fig.~\ref{fig:prhoslit}(b)
which shows the form of $\ln p(\rho)$ corresponding to the fully
periodic system at coexistence, for the same value of $L$. The peak to
valley separation in this plot is the same as the vapor to liquid peak
separation in the slit system at $\epsilon_{wf}=0.5$. It follows from
Eqs.(\ref{eq:Young})-(\ref{eq:gamdiff}) that the vanishing of the
liquid peak in $\ln p(\rho)$ marks the drying point $\cos(\theta)=-1$.

\begin{figure}[h]
\includegraphics[type=pdf,ext=.pdf,read=.pdf,width=0.94\columnwidth,clip=true]{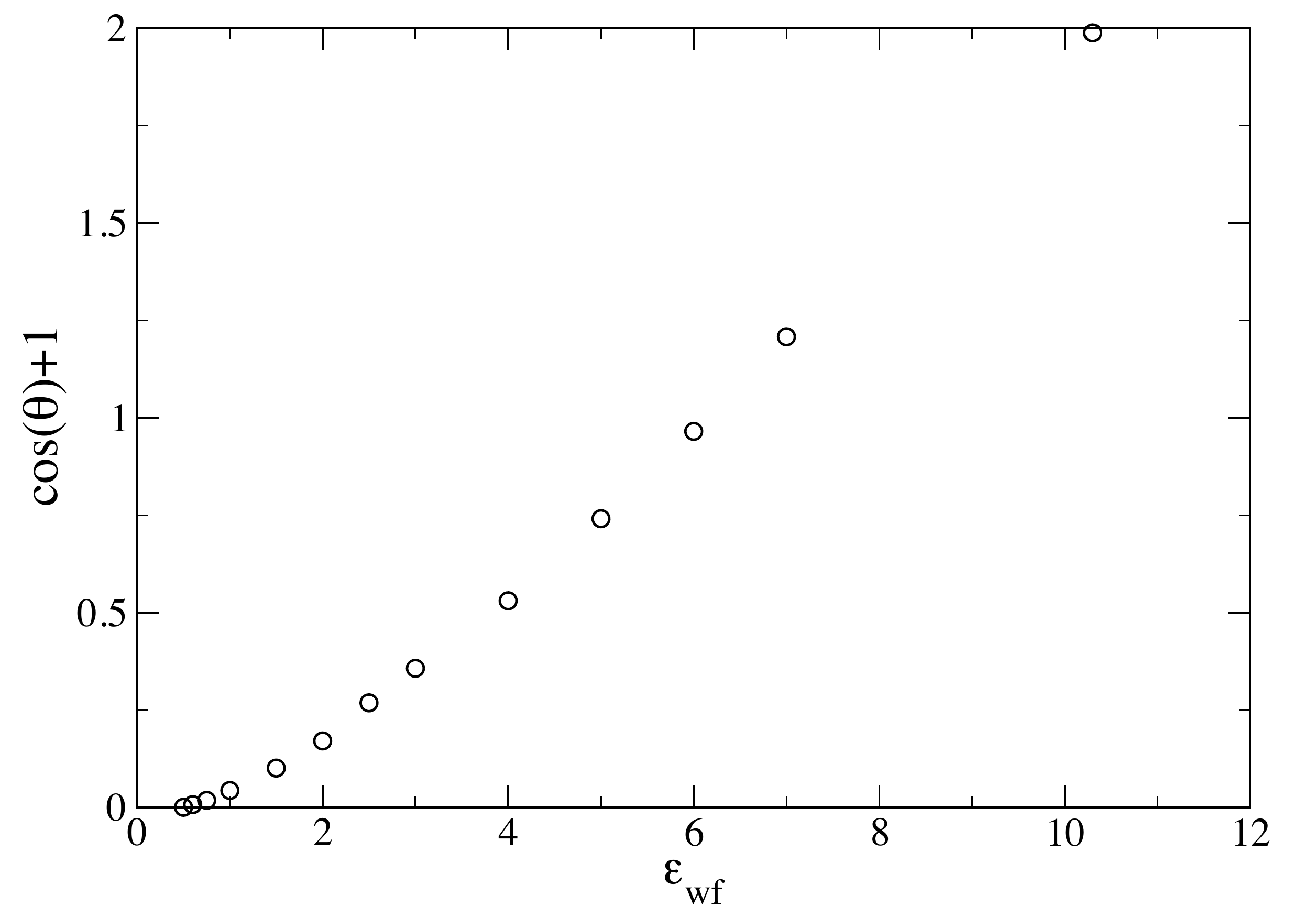}
\caption{Estimates of $\cos(\theta)+1$ as a function of
  $\epsilon_{wf}$ spanning the region from wetting $\cos(\theta)=1$ to
  drying $\cos(\theta)=-1$. Statistical uncertainties are smaller than symbol sizes.}

\label{fig:costheta}
\end{figure}

The behaviour of $\ln p(\rho)$ in the slit system as a function of
$\epsilon_{wf}$ reveals interesting qualitative differences between the
nature of the approach to complete wetting ($\cos(\theta)\to 1$), and
to complete drying ($\cos(\theta)\to -1$). In the former case, the
wall-vapor configurations remain metastable at the transition, as
evidenced by the presence of a vapor {\em peak} at the wetting wall
strength $\epsilon_{wf}=10.3$.  This signifies that the wetting
transition of water is a first order surface phase transition. By
contrast, on approaching the drying point, the liquid peak vanishes
smoothly into a plateau of constant probability. The distinction is borne
out by a plot of $\cos(\theta)+1$ versus $\epsilon_{wf}$ calculated from (\ref{eq:Young}) and shown in
Fig.~\ref{fig:costheta}. We find that $\cos(\theta)$ approaches unity
with a non-zero gradient denoting unambiguously a first order wetting transition, but appears to approach $-1$ tangentially. Such a scenario implies a continuous (critical) transition to drying.

\begin{figure}[h]
\includegraphics[type=pdf,ext=.pdf,read=.pdf,width=0.94\columnwidth,clip=true]{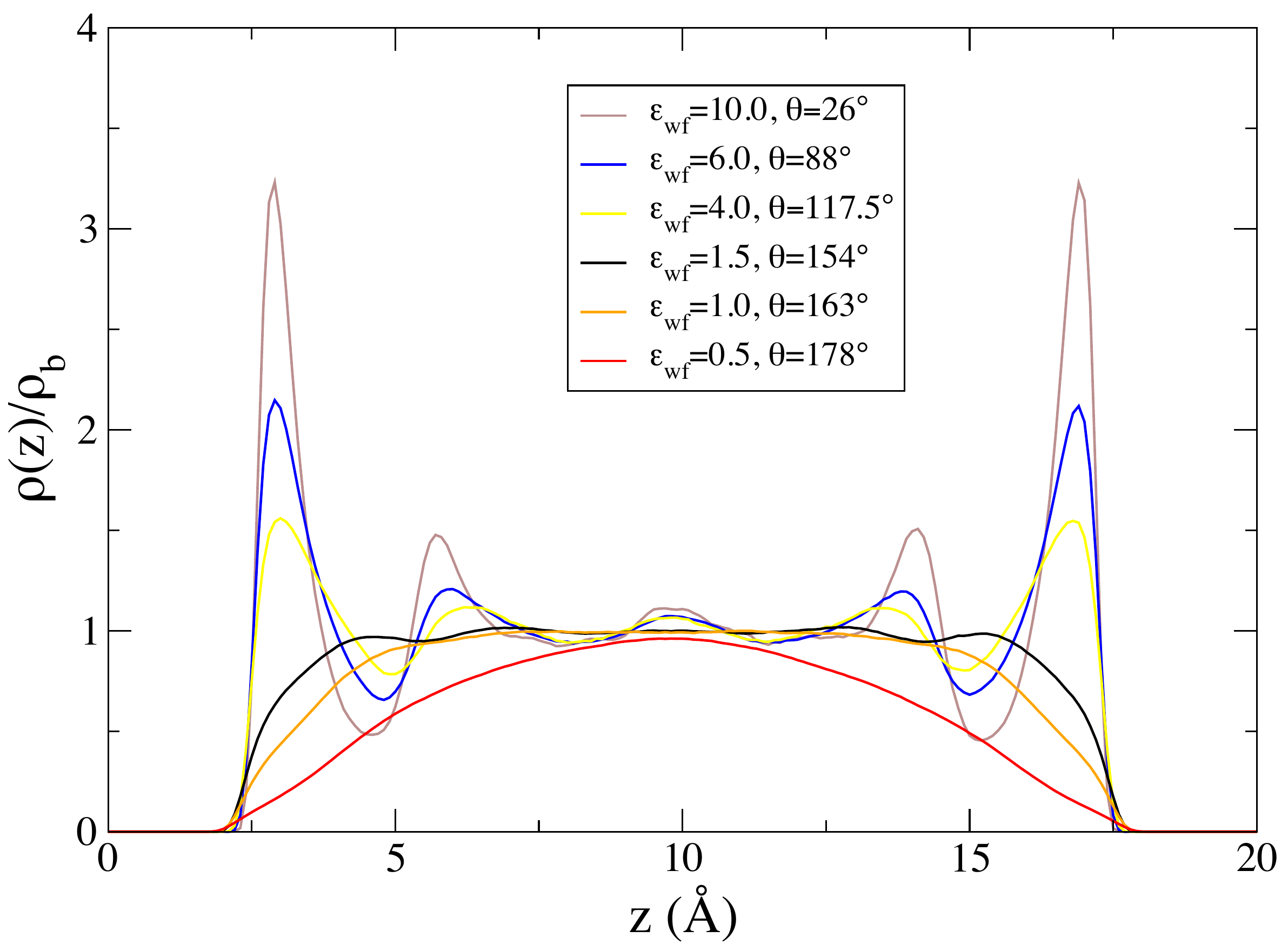}
\vspace{-2mm}
\caption{(Color online). Normalized number density profiles $\rho(z)$ for various values of $\epsilon_{wf}$ in the hydrophobic regime. $\rho_b$ is the bulk liquid density at $\mu_{cx}$.}
\label{fig:densprofs}
\end{figure}

The physical interpretation of the smooth erosion of the liquid peak
on the approach to drying is that the free energy barrier attaching
the liquid to the wall vanishes continuously. This allows the liquid
layer to detach and be replaced by vapor. In our slit system the
emergent liquid-vapor interface fluctuates freely (cf., the plateau in
$p(\rho)$ at $\epsilon_{wf}=0.5$) until the liquid slab thickness decreases
sufficiently that the system undergoes (capillary) evaporation.  This
interpretation is confirmed by measurements of the number density
profile $\rho(z)$ of the oxygen atoms in the hydrophobic regime
(Fig.~\ref{fig:densprofs}). For medium strength attractive wall-fluid
interactions which give rise to `neutral' hydrophobicity, i.e. $\theta\sim90^\circ$, the liquid density is high at the wall and packing
effects occur. The structure in the profiles decreases smoothly as
$\epsilon_{wf}$ is reduced and a low density vapor layer starts to appear near the
wall, the thickness of which appears to grow continuously as drying is
approached. The thickness of the vapor layer represents the order parameter for the drying transition \cite{Evans:1989aa}.

\begin{figure}[h]
\includegraphics[type=pdf,ext=.pdf,read=.pdf,width=0.94\columnwidth,clip=true]{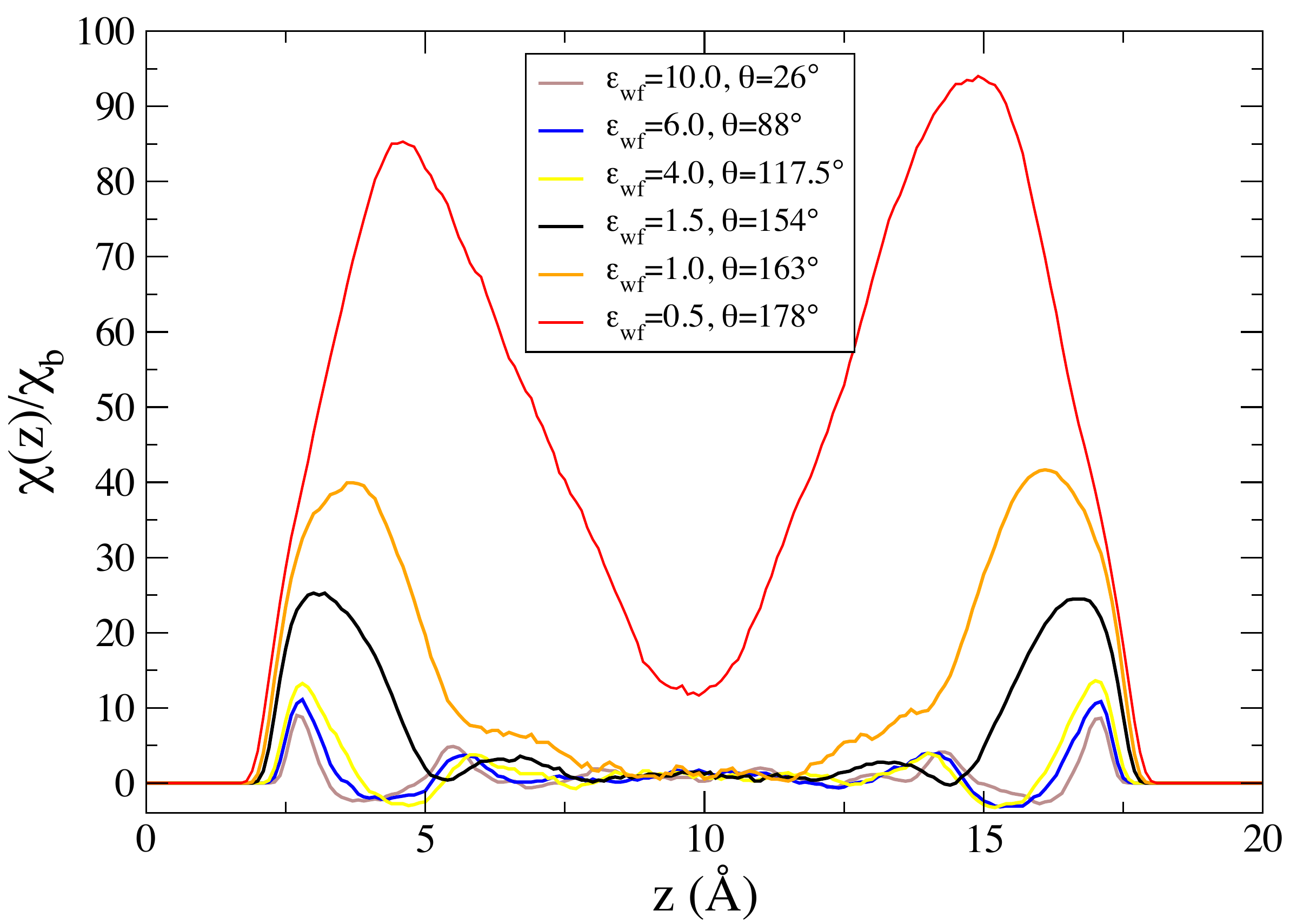}
\vspace{-2mm}
\caption{(Color online). The normalized local compressibility $\chi(x)/\chi_b$ for a selection of values of $\epsilon_{wf}$. These correspond to the density profiles of Fig.~(\protect\ref{fig:densprofs}).}
\label{fig:compnorm}
\end{figure}

The behaviour of $\rho(z)$ on the approach to drying reveals the
emergence of a vapor layer or density gap associated with hydrophobic
surfaces.  However, a much more sensitive and revealing measure of the
degree of hydrophobicity is the compressibility profile $\chi(z)$
defined in (\ref{eq:chi}), which provides a robust measure of local
density fluctuations close to the wall. Accurate estimates of this
quantity are readily obtained within the GCE, by exploiting histogram
reweighting to numerically differentiate the density profile
$\rho(z|\mu)$. The measured forms of $\chi(z)$, normalized with
respect to the bulk compressibility $\chi_b$ (i.e. the compressibility
far from the wall), are presented in Fig.~\ref{fig:compnorm}. These
show that close to the wall, $\chi(z)/\chi_b$ grows rapidly as
$\epsilon_{wf}$ is reduced towards the drying point, eventually
exceeding its bulk value by nearly two orders of
magnitude \footnote{Very close to the drying transition, the local
  fluctuations extend sufficiently far from each wall that the two
  walls of our slit start to interact. This is the reason that
  $\chi(z)$ does not decay to unity at $\epsilon_{wf}=0.5$. }. This
finding mirrors what is found in DFT calculations of the solvophobic
regime in a Lennard-Jones system \cite{Evans:2015aa}, and reflects the
development of a large transverse correlation length for density
fluctuations as $\cos(\theta)\to -1$~\cite{Evans:SM}. We note that
there is a large enhancement in the local compressibility even for
values of $\epsilon_{wf}$ which correspond to contact angles not much
greater than $\theta=90^\circ$ \footnote{Even in the hydrophilic
  regime $\theta<90^\circ$, $\chi(z)$ can exceed its bulk value significantly, and is
  seen to follow the oscillatory form of $\rho(z)$. However, this surface
  enhancement arises not due to criticality but reflects the fact that
  $\chi(z)$ is coupled to the growth of packing effects in $\rho(z)$.}. This
confirms the relevance of our findings for experimental studies on
real smooth hydrophobic surfaces such as SAMs where contact angles
have a maximum value of about $130^\circ$.

In summary, we have shown that smart sampling within the GCE is a powerful approach for accurately
characterising the hydrophobic regime including the approach to drying ($\theta\to 180^\circ$) in
water. The local compressibility profile $\chi(z)$ is the proper
statistical mechanics measure of local density fluctuations in fluids~\cite{Evans:SM}. When applied to water in
contact with a hydrophobic substrate, $\chi(z)$ provides a sensitive indicator of how the
microscopic structure near the substrate reflects the macroscopic contact angle $\theta$ --much more
so than the density profile $\rho(z)$ alone. In the drying limit the order parameter (i.e. the
thickness of the vapor layer) grows continuously but slowly with decreasing $\epsilon_{wf}$. In
contrast $\chi(z)$ grows rapidly and exceeds its bulk value by nearly two orders of magnitude over
the range of $\epsilon_{wf}$ explored, indicating the growth of a large transverse correlation
length~\cite{Evans:SM}. These findings point to drying in water at models of hydrophobic substrates
as being a surface critical phenomenon. Indeed separate simulation studies of drying in the
Lennard-Jones system reproduce the phenomenology seen in our studies of water whilst increased
system size permits the extraction of critical power law behaviour \cite{Evans:ab}. With these
insights, one can rationalize and explain the observations of enhanced fluctuations in previous
simulation studies of water near hydrophobic surfaces
\cite{Acharya:2010aa,Godawat:2011aa,Chandler:2007aa,
Patel:2010dz,Willard:2014aa,Kumar:2013kx,Mittal:2010aa}. We believe that these are attributable to
the proximity of a surface critical point i.e. the approach to a continuous drying transition, the
effects of which extend throughout the hydrophobic regime but were not recognised previously.

\acknowledgments

Some of the simulations described here were performed on the Bath HPC
Cluster. We thank R. Jack for helpful discussions.

%

\newpage
\end{document}